\documentclass [12pt,a4paper]{article}
\usepackage[cp1251]{inputenc}
\usepackage[english]{babel}
\usepackage[final]{graphicx}

\newcommand{\beq}[1]{\begin{equation} \label{#1} }
\newcommand{\eeq}   {\end{equation}}
\newcommand{\ds}{\displaystyle \mathstrut}
\newcommand{\Frac}[2]{\frac
{\textstyle\lefteqn{\phantom{{}_{\mathstrut}^{\mathstrut}}} #1}
{\textstyle\lefteqn{\phantom{{}_{\mathstrut}^{\mathstrut}}} #2}}

\newcommand{\av}[1]{\langle #1 \rangle}

\begin{document}
\title{Testing T Invariance in the Interaction of Slow Neutrons with Aligned Nuclei}
\author{A. L. Barabanov $^1$, A. G. Beda $^2$\\
{\normalsize\it $^1$ Kurchatov Institute, Moscow 123182, Russia}\\
{\normalsize\it $^2$ Institute for Theoretical and Experimental Physics,
Moscow 117259, Russia}}
\date{}
\maketitle

\begin{abstract}
The study of five-fold (P even, T odd) correlation in the interaction of slow polarized neutrons with aligned nuclei is a possible way of testing the time reversal invariance due to the expected enhancement of T violating effects in compound resonances. Possible nuclear targets are discussed which can be aligned both dynamically as well as by the "brute force" method at low temperature. A statistical estimation is performed of the five-fold correlation for low lying p wave compound resonances of the $^{121}$Sb, $^{123}$Sb and $^{127}$I nuclei. It is shown that a significant improvement can be achieved for the bound on the intensity of the fundamental parity conserving time violating (PCTV) interaction.
\end{abstract}
\bigskip

PACS: 24.80.+y Nuclear tests of fundamental interactions and symmetries

\section{Introduction}

The existence of T violating (TV) fundamental interactions beyond the Standard Model remains one of the most intriguing questions of modern physics. This is a reason to look for T odd angular correlations in nuclear reactions caused by possible TV interactions. Among the different reactions an interaction of slow neutrons with medium and heavy nuclei has an advantage. Indeed, the capture of a neutron by the nucleus results in the formation of narrow compound resonances. As was first noted in \cite{Bun82,Bun88}, the long life time of the resonances should lead to a large enhancement of T odd effects in the same way as occurs for P odd ones \cite{Sus80}. Studies of the enhanced P violating (PV) effects in neutron-nuclei interaction are described, e.g., in the review paper \cite{Mit01}.

So far evidence for violations of CP and T invariances (related by CPT symmetry) has only been found in K$^0$ decay. This lack of evidence means that the origin of the interaction responsible for these violations cannot be established. In particular, there is no consensus on whether or not TV interactions conserve P invariance. So hypothetical PVTV (P Violating, T Violating) and PCTV (P Conserving, T Violating) interactions have to be compared.

Starting the 1960s, a lot of attempts have been made to discover T violation in systems differing from K$^0$ mesons, specifically in nuclear reactions and decays. Only upper bounds have been set on the strength of both PCTV as well as PVTV interactions in the corresponding experiments. It is generally agreed (and there is a reason for it) that data on electric dipole moments (EDM) of neutron and atoms give the best bounds. Recall that the EDM of an elementary particle only arises at the simultaneous violation of P and T invariances.

Let
\beq{1.1}
\lambda_{PT}=\frac{\av{V_{PVTV}}_{s.p.}}{\av{V_{PV}}_{s.p.}},\quad
\lambda_T=\frac{\av{V_{PCTV}}_{s.p.}}{\av{V}_{s.p.}},
\eeq
be the ratios of typical single particle matrix elements of PVTV and PV interactions, as well as the ones of PCTV and strong interactions. Note that it is the practice to only compare PCTV interactions with strong interaction, $V$, because both are P even.

Current EDM data give the bound $\lambda_{PT} < 10^{-4}$ \cite{Her96}. The situation on the bound on PCTV interactions is more intricate. The PCTV forces alone cannot induce EDM because they are P even. However, PCTV forces together with weak PV interactions may involve EDM. Therefore, the bound on PCTV forces may be obtained from the EDM data but only in the framework of some model assumptions. During the 1990s there was confidence that the data on the EDM of neutrons and atoms put more strict bounds on PCTV interactions than direct searches for T odd, P even effects (e.g., in nuclear reactions). Summing the results of theoretical works devoted to relations between PCTV forces and EDM, a rather strict bound $\lambda_T < 10^{-8}$ was obtained in \cite{Her96}.

Recently this bound was reconsidered in \cite{Kur01}. More exactly, it was pointed out that the strict limitation on $\lambda_T$ from EDM data arises in the framework of a certain assumption ("Scenario A"). In fact, there are no real fundamental reasons to follow this assumption. In the framework of an alternative approach ("Scenario B") the EDM data give no definite value for $\lambda_T$. In this case one should use only a relatively weak bound $\lambda_T < 10^{-4}$ resulting from direct searches for T non-invariant, P even effects.

It was noted in \cite{Gou02,Bar03} that this re-estimation of the current bound on PCTV interactions favours resuming direct tests of T invariance by studies of P even, T odd effects in nuclear reactions and decays, specifically, in the interaction of slow neutrons with medium and heavy nuclei.

Recall that in nuclear reactions T invariance results in equality of the amplitudes (elements of S matrix) of direct, $S(\alpha\to\alpha')$, and inverse, $S(\alpha'\to\alpha)$, transitions. In particular, in the neutron-nucleus interaction one usually labels the scattering states by relative orbital momentum $l$, total angular momentum $j$ of the neutron (${\bf j}={\bf l}+{\bf s}$, where $s=1/2$ is the neutron spin) and total angular momentum $J$ of the neutron and target nucleus (${\bf J}={\bf j}+{\bf I}$, where $I$ is the nucleus spin). Then one uses the amplitudes $S_J(lj\to l'j')$ implying that $l$ and $j$ can change at scattering, while $J$ is conserved. Therefore, possible PCTV interaction results in a difference between $S_J(lj\to l'j')$ and $S_J(l'j'\to lj)$, where $l$ and $l'$ are of the same parity.

This work deals with the possibility of searching for PCTV interactions by studying P conserving, T violating five-fold correlation in the total cross section of the interaction of polarized slow neutrons with spin-aligned nuclei. This P even, T odd five-fold correlation $({\bf s}[{\bf k}\times{\bf I}])({\bf k}{\bf I})$, where ${\bf s}$, ${\bf I}$ and ${\bf k}$ are the neutron and nucleus spins and neutron momentum, respectively, was first considered in \cite{Bary83,Bar86,Kab86}. It only arises when the difference of the amplitudes $S_J(lj\to l'j')$ and $S_J(l'j'\to lj)$ is non-zero.

Five-fold correlation may be used to test T invariance (provided space parity is conserved) in the interaction of any polarized particles with spin-aligned targets. It must be emphasized that, first, this method is a genuine null test of T invariance and, second, it is free from masking effects related to final state interaction \cite{Con93,Con95}. We concentrate on the interaction of slow neutrons with medium and heavy nuclei by reason of the large enhancement pointed out above.

In practice, the studies are hindered to a great extent by the absence of appropriate nuclear targets. One needs nuclei which, first, can be aligned (with spin $I\ge 1$) and, second, have p wave resonances in the low energy region. Last few years, due to studies of P violation in the interaction of slow neutrons with nuclei, a list of known p wave resonances for many nuclei was considerably expanded. Among the targets discussed in \cite{Mit01}, the nuclei $^{121}$Sb, $^{123}$Sb and $^{127}$I are of special interest. Below (see Section 8) we demonstrate real possibilities of aligning them.

We begin this paper by discussion of various aspects of the interaction of slow polarized neutrons with aligned nuclei (Sections 2-5). Then, in Section 6, a new method is proposed for statistical estimation of the magnitudes of five-fold correlation (as well as of the deformation effect) in given p wave resonances. The T non-invariant, P even effect is proportional to the mean matrix element $v_T$ of PCTV interaction. Its value is considered in Section 7. Prospects of aligning the nuclei with known p wave resonances are studied in Section 8. In Section 9 in the framework of the advanced statistical approach, we perform numerical estimations of the discussed effects for known p wave resonances of $^{121}$Sb, $^{123}$Sb and $^{127}$I nuclei. It is shown that the results of the measurements of PV correlation \cite{Mit01} may be used to estimate the feasibility of the proposed measurement of five-fold correlation for the same nuclei.

Our conclusion is that the current bound on PCTV interaction could be improved (at best, PCTV effect would be found) even at the measurement conditions realized for PV studies \cite{Mit01}. New high intensity spallation neutron sources -- SNS (USA) and JSS (Japan) -- would provide better possibilities to reach the same goal.

\section{Correlations in the interaction of polarized neutrons with aligned nuclei}

Generally, the total cross section of neutron-nucleus interaction depends on the relative orientation of the unit vectors ${\bf n}_s$, ${\bf n}_I$ and ${\bf n}_k$ along the neutron polarization axis, nuclei alignment axis and neutron momentum, respectively. We use the following definitions,
\beq{2.1}
p_1(s)=\frac{\av{\sigma}}{s},\qquad
p_2(I)=\frac{3\av{m^2}-I(I+1)}{I(2I-1)},
\eeq
for neutron polarization ($\sigma$ is the projection of spin $s=1/2$ on the axis ${\bf n}_s$) and nuclei alignment ($m$ is the projection of spin $I$ on the axis ${\bf n}_I$), respectively. We consider the situation of pure alignment when the populations of nuclear substates with projections $m$ and $-m$ are equal (as the result $\av{m}=0$ and nuclei polarization $p_1(I)=\av{m}/I$ is equal to zero). Note that only nuclei with spin $I\ge 1$ can be aligned.

The neutron-nucleus interaction at low energies is predominantly s wave and, therefore, is described by the sole amplitude $S_J(0\frac{1}{2}\to 0\frac{1}{2})$. At first glance it renders the test of T invariance impossible because the initial and final states coincide. In fact, in the low lying p wave resonances contributions of partial p1/2 and p3/2 waves can be comparable with the contribution of s wave. Because of this, in \cite{Bar86} it was proposed to look for five-fold correlation just in p wave resonances and, hence, an explicit expression was obtained for the effect in terms of the S matrix on the assumption that only s1/2, p1/2 and p3/2 partial waves are of importance. Namely, it was shown that the total cross section of the neutron-nucleus interaction is of the form
\beq{2.2}
\begin{array}{l}
\sigma_{tot}=\sigma_0+p_1(s)({\bf n}_s{\bf n}_k)\Delta\sigma^{(1)}_{PV}+
p_2(I)(3({\bf n}_k{\bf n}_I)^2-1)\Delta\sigma_D+{}
\\[\bigskipamount]
\phantom{\sigma_{tot}=\sigma_0}+
p_1(s)p_2(I)(3({\bf n}_s{\bf n}_I)({\bf n}_k{\bf n}_I)-
({\bf n}_s{\bf n}_k))\Delta\sigma^{(2)}_{PV}+{}
\\[\bigskipamount]
\phantom{\sigma_{tot}=\sigma_0}+
p_1(s)p_2(I)({\bf n}_s[{\bf n}_k\times {\bf n}_I])
({\bf n}_k{\bf n}_I)\Delta\sigma_{PCTV},
\end{array}
\eeq
were $\sigma_0$ is the total cross section for unpolarized neutrons and non-oriented
nuclei
\beq{2.3}
\sigma_0=\frac{2\pi}{k^2}\sum_Jg_J\sum_{lj}\left(
1-{\rm Re}\,S_J(lj\to lj)\right),
\eeq
and
\beq{2.4}
\Delta\sigma_D=\frac{\pi}{k^2}{\ds\sum_J}
\left(h_J\,{\rm Re}\left(
S_J(1\frac{1}{2}\to 1\frac{3}{2})+
S_J(1\frac{3}{2}\to 1\frac{1}{2})\right)+
e_J\left(1-{\rm Re}\,S_J(1\frac{3}{2}\to 1\frac{3}{2})\right)\right),
\eeq
\beq{2.5}
\Delta\sigma^{(1)}_{PV}=\frac{2\pi}{k^2}\sum_J
g_J\,{\rm Re}\left(S_J(0\frac{1}{2}\to 1\frac{1}{2})+
S_J(1\frac{1}{2}\to 0\frac{1}{2})\right),
\eeq
\beq{2.6}
\Delta\sigma^{(2)}_{PV}=\frac{\pi}{k^2}\sum_J
h_J\,{\rm Re}\left(S_J(0\frac{1}{2}\to 1\frac{3}{2})+
S_J(1\frac{3}{2}\to 0\frac{1}{2})\right),
\eeq
\beq{2.7}
\Delta\sigma_{PCTV}=\frac{3\pi}{k^2}\sum_J
h_J\,{\rm Im}\left(S_J(1\frac{1}{2}\to 1\frac{3}{2})-
S_J(1\frac{3}{2}\to 1\frac{1}{2})\right).
\eeq
The numerical factors $g_J$, $h_J$ and $e_J$ are given by the following expressions
\beq{2.8}
g_J=\frac{2J+1}{2(2I+1)},
\eeq
\beq{2.9}
h_{I-1/2}=\frac{I}{2I+1}\sqrt{\frac{2I-1}{I+1}},\quad
h_{I+1/2}=-\frac{2I-1}{2I+1}\sqrt{\frac{I}{2I+3}},\quad
h_{I\pm 3/2}=0,
\eeq
\beq{2.10}
\begin{array}{l}
e_{I-3/2}=-\Frac{I-1}{2I+1},\quad
e_{I-1/2}=\Frac{I(I-2)}{(2I+1)(I+1)},
\\[\bigskipamount]
\phantom{e_{I-3/2}={}}
e_{I+1/2}=\Frac{(2I-1)(I+3)}{(2I+1)(2I+3)},\quad
e_{I+3/2}=-\Frac{(2I-1)I(I+2)}{(2I+1)(I+1)(2I+3)}.
\end{array}
\eeq

Here the factor (\ref{2.7}) at five-fold correlation is proportional to the difference between the amplitudes $S_J(1\frac{1}{2}\to 1\frac{3}{2})$ and $S_J(1\frac{3}{2}\to 1\frac{1}{2})$. Later an additional contribution from the mixing of s and d waves was considered \cite{Gou90, Bun90, Gud91}, which is proportional to the difference between the amplitudes $S_J(0\frac{1}{2}\to 2j)$ and $S_J(2j\to 0\frac{1}{2})$, where $j=3/2$ and $5/2$. This is of special importance for deformed nuclei where s wave resonances may have appreciable d wave partial widths \cite{Huf98}. The prospects for searching for five-fold correlation in s wave resonances (with energies up to 500 eV) of deformed nucleus $^{165}$Ho were recently discussed in \cite{Huf02}.

Note that searching for five-fold correlation in the interaction of fast polarized neutrons (with energy close to 6 MeV) with aligned nuclei $^{165}$Ho has been performed in \cite{Huf97}. A strict bound on the PCTV interaction was obtained. However, a similar study on slow neutrons has not been carried out so far.

\section{Neutron-nucleus interaction in p wave resonance}

All angular correlations in (\ref{2.2}) are pronounced in p wave resonances because they are induced by p waves (see (\ref{2.4})--(\ref{2.7})). Let p wave resonances with spin $J$ be labelled by the index $\mu$. Then we assume that $E_{\mu}^J$ is the position of the $\mu$th resonance, $\Gamma_{\mu}^J$ is the total width, $g_{n\mu}^J(1j)$ is the real amplitude of the partial neutron width $\Gamma_{n\mu}^J(1j)=(g_{n\mu}^J(1j))^2$,
and $\Gamma_{n\mu}^J=\Gamma_{n\mu}^J(1\frac{1}{2})+\Gamma_{n\mu}^J(1\frac{3}{2})$ is the total neutron width. Close to the $\mu$th resonance the p wave scattering amplitude is of the form
\beq{2.11}
S_J(1j\to 1j')=e^{2i\delta_p}
\left(\delta_{jj'}-
i\frac{g_{n\mu}^J(1j)g_{n\mu}^J(1j')}{E-E_{\mu}^J+i\Gamma_{\mu}^J/2}\right)+
\Delta S_J^{PCTV}(1j\to 1j').
\eeq
Here $\delta_p\simeq (kR)^3/3$ is the potential scattering phase for the p wave on the nucleus with radius $R$. Below we neglect it. The small additive $\Delta S_J^{PCTV}$ resulting from the hypothetical PCTV interaction is considered in details later (for the moment we omit it).

Close to the $\mu$th resonance one obtains p wave contribution to the total cross section $\sigma_0$ by substitution of (\ref{2.11}) into (\ref{2.3}). The result is described by the Breit--Wigner formula
\beq{2.12}
\sigma_p(E)=\frac{\pi}{k^2}\,
\frac{g_J\,\Gamma_{n\mu}^J\Gamma_{\mu}^J}
{(E-E_{\mu}^J)^2+(\Gamma_{\mu}^J/2)^2}.
\eeq
Since the neutron width of a p wave resonance is small, the cross section $\sigma_p$ adds little in comparison to the s wave contribution $\sigma_s$ into $\sigma_0$ (even at $E=E_{\mu}^J$).

For this reason a small number of p wave resonances is known. It has already been pointed out that over the last few years a lot of new p wave resonances were found due to PV studies on the neutron-nucleus interaction. Note that the statistical properties of p wave resonances have been investigated much less  than those of s wave ones and, therefore, are of separate interest. In particular, there is no information about the distribution and mean values of partial neutron amplitudes $g_{n\mu}^J(1\frac{1}{2})$ and $g_{n\mu}^J(1\frac{3}{2})$ (see, e.g., \cite{Bar87}).

\section{Polarized neutrons and non-aligned nuclei: PV effect}

Let target nuclei be not aligned, $p_2(I)=0$, while neutrons are polarized. Then on the right-hand side of Eq.(\ref{2.2}) apart from $\sigma_0$ there is the sole term which is proportional to P odd correlation $({\bf n}_s{\bf n}_k)$ of neutron spin and momentum. The evident expression (\ref{2.5}) for $\Delta\sigma^{(1)}_{PV}$ clearly demonstrates that this correlation owes its existence to parity violating forces resulting in transitions from s wave to p wave and vice versa. Strong enhancement of this PV correlation in p wave resonances was first predicted in \cite{Sus80} in a model of mixing of s and p wave resonances. Note that PV forces mix s and p wave resonances with equal spins.

Let s wave resonances with spin $J$ be labelled by the index $\nu$, and $V^{PV}_{\nu\mu}$ is the real matrix element of the PV interaction between the wave functions of the $\nu$th and $\mu$th compound states. Thus, in accordance with \cite{Sus80} (see also \cite{Mit01}), close to the $\mu$th resonance the main contribution to the S matrix elements in (\ref{2.5}) (as well as in (\ref{2.6})) has the form
\beq{2.16}
S_J(0\frac{1}{2}\to 1j)=S_J(1j\to 0\frac{1}{2})=
-i\frac{g_{n\mu}^J(1j)}
{E-E_{\mu}^J+i\Gamma_{\mu}^J/2}
\sum_{\nu}\frac{V_{\nu\mu}^{PV}g_{n\nu}^J(0\frac{1}{2})}{E_{\mu}^J-E_{\nu}^J},
\eeq
where $E_{\nu}^J$ is the position of $\nu$th resonance, and $g_{n\nu}^J(0\frac{1}{2})$ is the real amplitude of the neutron width $\Gamma_{n\nu}^J=(g_{n\nu}^J(0\frac{1}{2}))^2$ of the same resonance. Substitution of (\ref{2.16}) into (\ref{2.5}) gives
\beq{2.17}
\Delta\sigma^{(1)}_{PV}(E)=\frac{\pi}{k^2}\,
\frac{g_J\,g_{n\mu}^J(1\frac{1}{2})\Gamma_{\mu}^J}
{(E-E_{\mu}^J)^2+(\Gamma_{\mu}^J/2)^2}
\sum_{\nu}\frac{2V_{\nu\mu}^{PV}g_{n\nu}^J(0\frac{1}{2})}{E_{\nu}^J-E_{\mu}^J}.
\eeq
It is clear that close to p wave resonance this factor has the same (resonance) dependence on the neutron energy $E$ as the p wave cross section (\ref{2.12}).

In experiment, one measures the asymmetry of neutron transmission trough the target for two opposite directions of neutron spin (along and opposite the neutron momentum)
\beq{2.18}
\alpha_{PV}=\frac{N_+-N_-}{N_++N_-},
\eeq
where $N_+$ and $N_-$ are the numbers of neutrons travelling through the target at ${\bf n}_s={\bf n}_k$ and ${\bf n}_s=-{\bf n}_k$, respectively. Let $n$ be the number of nuclei in the unit volume of the target, and $d$ the target thickness. Then taking into account that $\Delta\sigma^{(1)}_{PV}\ll\sigma_0$, one gets
\beq{2.19}
\alpha_{PV}=-ndp_1(s)\Delta\sigma^{(1)}_{PV}.
\eeq
Thus, in accordance with (\ref{2.17}) the measured PV asymmetry $\alpha_{PV}$ is enhanced in the $\mu$th p wave resonance (and reaches a maximum at $E=E_{\mu}^J$). This enhancement is said to be resonance \cite{Bun81}.

To analyze the results, one extracts the factor (\ref{2.17}) from $\alpha_{PV}$. Then, to eliminate the dependence on the neutron energy, one usually divides $\Delta\sigma^{(1)}_{PV}$ into the p wave cross section (\ref{2.12}) and deals with the quantity
\beq{2.20}
p_P\equiv\frac{\Delta\sigma^{(1)}_{PV}}{\sigma_p}=
\frac{\Gamma_{n\mu}^J(1\frac{1}{2})}{\Gamma_{n\mu}^J}\sum_{\nu}
\frac{2V_{\nu\mu}^{PV}}{E_{\nu}^J-E_{\mu}^J}\,
\frac{g_{n\nu}^J(0\frac{1}{2})}{g_{n\mu}^J(1\frac{1}{2})}.
\eeq
It includes the enhancement factor $g_{n\nu}^J(0\frac{1}{2})/g_{n\mu}^J(1\frac{1}{2})\sim (kR)^{-1}\sim 10^2$-$10^3$ which is said to be kinematical (or structural) \cite{Sus80,Bun81}. It arises due to the suppression of the amplitudes of neutron widths of p wave resonances with respect to the ones of s wave resonances by the factor $kR\sim 10^{-2}$-$10^{-3}$.

At present the quantities $p_P$ are measured for a lot of p wave resonances for more than two tens of nuclei \cite{Mit01}. The results are treated on the assumption that for any nucleus the matrix elements $V_{\nu\mu}^{PV}$ are Gaussian random quantities with a root-mean square $v_P$. The extracted values $v_P$ are close to 1~meV for all investigated nuclei.

\section{Aligned nuclei: deformation effect, PV effect and TV effect}

When target nuclei are aligned, three additional terms appear in the right-hand side of Eq.(\ref{2.2}). Any of them
can be measured.

The first term is proportional to the correlation $(3({\bf n}_k{\bf n}_I)^2-1)$ of neutron momentum and nucleus spin. No neutron polarization is required. Using the same notations as in \cite{Huf98} we designate this term as the "deformation term"\footnote{It corresponds to the visual interpretation of this correlation. Indeed, let us assume that the nucleus is a prolate spheroid with spin ${\bf I}$ along the major axis. Then different nuclear cross sections correspond to the longitudinal and transverse orientation of ${\bf I}$ with respect to ${\bf k}$.}. Close to the $\mu$th resonance, one gets the explicit expression for $\Delta\sigma_D$ substituting (\ref{2.11}) into (\ref{2.4}),
\beq{2.13}
\Delta\sigma_D(E)=\frac{\pi}{k^2}
\left(-h_Jg_{n\mu}^J(1\frac{1}{2})g_{n\mu}^J(1\frac{3}{2})+
\frac{e_J}{2}\Gamma_{n\mu}^J(1\frac{3}{2})\right)
\frac{\Gamma_{\mu}^J}
{(E-E_{\mu}^J)^2+(\Gamma_{\mu}^J/2)^2}.
\eeq
We point to its resonance dependence on the neutron energy.

Deformation effect determines the asymmetry of neutron transmission through the target with alignment axis directed along, ${\bf n}_I\|{\bf n}_k$, and transverse to, ${\bf n}_I\perp {\bf n}_k$, the neutron momentum, i.e.
\beq{2.14}
\alpha_D=\frac{N_{\|}-N_{\perp}}{N_{\|}+N_{\perp}}=
-ndp_2(I)\frac{3\Delta\sigma_D}{2},
\eeq
where one assumes that $\Delta\sigma_D\ll\sigma_0$. By analogy with (\ref{2.20}) in any p wave resonance one can introduce the quantity
\beq{2.15.2}
p_D\equiv\frac{\Delta\sigma_D}{\sigma_p}=
-\frac{h_J}{g_J}\,
\frac{g^J_{n\mu}(1\frac{1}{2})g^J_{n\mu}(1\frac{3}{2})}{\Gamma^J_{n\mu}}
+\frac{e_J}{2g_J}\,
\frac{\Gamma^J_{n\mu}(1\frac{3}{2})}{\Gamma^J_{n\mu}},
\eeq
which describes the magnitude of the deformation effect only in this resonance.

The second additional term in the right-hand side of (\ref{2.2}) is proportional to $\Delta\sigma^{(2)}_{PV}$ and results from P violation. Using (\ref{2.6}) and (\ref{2.16}) we get in the $\mu$th p wave resonance
\beq{2.21}
\Delta\sigma^{(2)}_{PV}(E)=\frac{\pi}{k^2}\,
\frac{h_Jg_{n\mu}^J(1\frac{3}{2})\Gamma_{\mu}^J}
{(E-E_{\mu}^J)^2+(\Gamma_{\mu}^J/2)^2}
\sum_{\nu}\frac{V_{\nu\mu}^{PV}g_{n\nu}^J(0\frac{1}{2})}{E_{\nu}^J-E_{\mu}^J}.
\eeq
It is apparent that the measurement of the asymmetry (\ref{2.18}) at different angles between the alignment axis ${\bf n}_I$ and neutron momentum ${\bf n}_k$ in line with the measurement of the deformation effect (\ref{2.14}), (\ref{2.15.2}) would provide information on the division of the neutron width $\Gamma_{n\mu}^J$ into partial widths $\Gamma_{n\mu}^J(1\frac{1}{2})$ and $\Gamma_{n\mu}^J(1\frac{3}{2})$, and, conceivably, on spin $J$ of the resonance.

However, the main subject of this paper is the five-fold correlation, i.e. the last term on the right-hand side of (\ref{2.2}). To find the dependence of the effect on the parameters of p wave resonance we use the model proposed in \cite{Bun88,Bun90}. This implies mixing of the p wave resonances by the PCTV interaction. Let $V_{\mu'\mu}^{PCTV}$ is the real matrix element of the PCTV interaction between the wave functions of the $\mu'$th and $\mu$th compound states. Then in the $\mu$th p wave resonance the main contribution to the PCTV additive in (\ref{2.11}) is of the form
\beq{2.22}
\begin{array}{l}
\Delta S_J^{PCTV}(1\Frac{1}{2}\to 1\Frac{3}{2})=
-\Delta S_J^{PCTV}(1\Frac{3}{2}\to 1\Frac{1}{2})={}
\\[\bigskipamount]
\phantom{\Delta S_J^{PCTV}}=
\Frac{1}{E-E_{\mu}^J+i\Gamma_{\mu}^J/2}
{\ds\sum_{\mu'\ne\mu}}\Frac{V_{\mu'\mu}^{PCTV}
\left(g_{n\mu}^J(1\frac{1}{2})g_{n\mu'}^J(1\frac{3}{2})-
g_{n\mu}^J(1\frac{3}{2})g_{n\mu'}^J(1\frac{1}{2})\right)}
{E_{\mu}^J-E_{\mu'}^J}.
\end{array}
\eeq
Thus (\ref{2.7}) transforms into
\beq{2.23}
\Delta\sigma_{PCTV}=\frac{\pi}{k^2}\,
\frac{h_J\Gamma_{\mu}^J}
{(E-E_{\mu}^J)^2+(\Gamma_{\mu}^J/2)^2}
\sum_{\mu'\ne\mu}\frac{3V_{\mu'\mu}^{PCTV}
\left(g_{n\mu'}^J(1\frac{1}{2})g_{n\mu}^J(1\frac{3}{2})-
g_{n\mu'}^J(1\frac{3}{2})g_{n\mu}^J(1\frac{1}{2})\right)}
{E_{\mu}^J-E_{\mu'}^J}.
\eeq
Note that this factor has the same resonance dependence on $E$ as the p wave cross section (\ref{2.12}).

To test T invariance one should measure the asymmetry
\beq{2.24}
\alpha_{PCTV}=
\frac{N_{\uparrow\uparrow}-N_{\downarrow\uparrow}}
{N_{\uparrow\uparrow}+N_{\downarrow\uparrow}},
\eeq
where $N_{\uparrow\uparrow}$ and $N_{\downarrow\uparrow}$ are the numbers of neutrons travelling through the target with polarization ${\bf n}_s\uparrow\uparrow [{\bf n}_k\times {\bf n}_I]$ and ${\bf n}_s\downarrow\uparrow [{\bf n}_k\times {\bf n}_I]$, respectively. Then
\beq{2.25}
\alpha_{PCTV}=-ndp_1(s)p_2(I)\sin\theta\cos\theta\Delta\sigma_{PCTV},
\eeq
where $\theta$ is the angle between ${\bf n}_I$ and ${\bf n}_k$. The asymmetry (\ref{2.25}) as function of $\theta$ has a maximum at $\theta=45^0$. Because the resonance enhancement takes place for $\Delta\sigma_{PCTV}$ and $\alpha_{PCTV}$ as well as for PV effects, it is natural to deal with the ratio
\beq{2.26}
p_T\equiv\frac{\Delta\sigma_{PCTV}}{\sigma_p}=
\frac{h_J}{g_J}\sum_{\mu'\ne\mu}
\frac{3V_{\mu'\mu}^{PCTV}}{E_{\mu}^J-E_{\mu'}^J}\,
\frac{g_{n\mu'}^J(1\frac{1}{2})g_{n\mu}^J(1\frac{3}{2})-
g_{n\mu'}^J(1\frac{3}{2})g_{n\mu}^J(1\frac{1}{2})}
{\Gamma_{n\mu}^J},
\eeq
which is analogous to (\ref{2.20}).

It should be stressed that the proposed measurement of the asymmetry (\ref{2.24})--(\ref{2.26}), which is sensitive to the possible violation of T invariance, is similar in procedure to the well developed (see \cite{Mit01}) measurement of the asymmetry (\ref{2.18})--(\ref{2.20}), which is sensitive to P violation. The quantity $p_T$ is to be measured in the same p wave resonances as the quantity $p_P$ was measured. There are no grounds to expect any correlations between $p_P$ and $p_T$. In fact, these quantities owe their existence to different interactions, PV and PCTV, respectively. Moreover, PV asymmetry results from mixing of p wave resonance with s wave resonances, while PCTV asymmetry from mixing of p wave resonance with other p wave resonances.

Nevertheless, the measurements \cite{Mit01} of $p_P$ are of great importance for the following reason. Let us assume that $p_T$ is measured in the same conditions (the same target and neutron flux) as the quantity $p_P$ was measured. We take into account that the statistical accuracy $\Delta p_P$ in the $\mu$th p wave resonance (as well as $\Delta p_T$) is mainly determined by the strength of this resonance and the s wave background. Therefore, one can expect that the error $\Delta p_T$ will be of the same magnitude as $\Delta p_P$ (to be more precise, slightly higher because on the right-hand side of (\ref{2.25}) at $\theta=45^0$ there is an attenuation factor, $p_2(I)/2$). Thus, comparing an evaluated value for $p_T$ in definite p wave resonance with the achieved accuracy $\Delta p_P$ in the same resonance, one can estimate the feasibility of the proposed study of five-fold correlation.

\section{Statistical estimation of TV and deformation effects}

In a fixed p wave resonance with spin $J$, the quantity $p_T$ (\ref{2.26}) cannot be predicted because, first, matrix elements $V_{\mu'\mu}^{PCTV}$ and, second, amplitudes $g_{n\mu}^J(1j)$ are unknown. Moreover, spins of p wave resonances where PV effects were measured are also unknown. For any p wave resonance we know only its position $E_{\mu}$ and product $g_J\Gamma_{n\mu}^J$ of statistical factor $g_J$ into neutron width $\Gamma_{n\mu}^J$.

Nevertheless, one can estimate a root-mean square
\beq{3.3}
\bar p_T=\sqrt{\av{p^2_T}}
\eeq
for the PCTV effect in the $\mu$th resonance with spin $J$. Indeed, let us assume that the matrix elements $V_{\mu'\mu}^{PCTV}$ are Gaussian random quantities with
\beq{3.1}
\av{V_{\mu'\mu}^{PCTV}}=0,\quad
\sqrt{\av{\left(V_{\mu'\mu}^{PCTV}\right)^2}}=v_T.
\eeq
The neutron amplitudes of p wave resonances have similar properties,
\beq{3.2}
\av{g_{n\mu}^J(1j)}=0,\quad
\sqrt{\av{\left(g_{n\mu}^J(1j)\right)^2}}=\frac{\Gamma_{n\mu}^J}{2},
\eeq
at the same time the amplitudes $g_{n\mu}^J(1\frac{1}{2})$ and $g_{n\mu}^J(1\frac{3}{2})$ are independent random quantities. Then we obtain from (\ref{2.26}), (\ref{3.1}) and (\ref{3.2})
\beq{3.4}
\bar p_T=
\frac{3|h_J|v_T}{\sqrt{2}\,g_J}
\left(\sum_{\mu'\ne\mu}\frac{1}{(E_{\mu}^J-E_{\mu'}^J)^2}\,
\frac{\Gamma_{n\mu'}^J}{\Gamma_{n\mu}^J}\right)^{1/2},
\eeq
where one sums over resonances $\mu'$ with the same spin $J$.

As a rule, recall, the spins of p wave resonances are unknown. Only for the resonances where PV effects were observed can one be sure that $J=I\pm 1/2$. An absence of the PV effect in a p wave resonance may results both from $J=I\pm 3/2$ as well as from negligibly small mixing of this resonance (with $J=I\pm 1/2$) with s wave resonances.

Based on the known formula for density of states with spin $J$, let us assume that the probability $w(J)$ for the p wave resonance to have spin $J$ is proportional to $(2J+1)$. Then, if $I$ is the spin of target nuclei (and $I\ge 3/2$), we get
\beq{3.5}
w(J)=\frac{g_J}{2},\quad
\sum_{J=I\pm 1/2, 3/2}w(J)=1.
\eeq
It is easily seen that a p wave resonance has spin $J=I\pm 1/2$ with probability 1/2 and spin $J=I\pm 3/2$ with the same probability 1/2. Below we perform an estimation of $\bar p_T$ in the $\mu$th resonance assuming that its spin is equal to $J=I-1/2$ or $I+1/2$. Thus, approximately one half of our (non-zero) estimates are improper; indeed, approximately one half of the known p wave resonances have spin $J=I\pm 3/2$ and, therefore, PCTV effect for these resonances is equal to zero.

So we let the $\mu$th resonance have spin $J=I-1/2$ or $I+1/2$. To evaluate $\av{p^2_T}$, one squares both sides of (\ref{2.26}), then averages them using (\ref{3.1}) and (\ref{3.2}), which yields the sum over the resonances $\mu'$ with the same spin $J$. Since any of the known p wave resonances can have spin $J$ with probability $w(J)$, we perform the summation over all known p wave resonances taking their contribution with weight factor $w(J)$. Thus using (\ref{3.5}), we replace Eq. (\ref{3.4}) by
\beq{3.6}
\bar p_T=
\frac{3|h_J|v_T}{2\sqrt{g_J}}
\left(\sum_{\mu'\ne\mu}\frac{1}{(E_{\mu}-E_{\mu'})^2}\,
\frac{\Gamma_{n\mu'}^J}{\Gamma_{n\mu}^J}\right)^{1/2},
\eeq
where one sums over all known p wave resonances. Note that the ratio of the neutron widths $\Gamma_{n\mu'}^J$ and $\Gamma_{n\mu}^J$ entering (\ref{3.6}) is equal to the ratio of known quantities $g_J\Gamma_{n\mu'}^J$ and $g_J\Gamma_{n\mu}^J$. The common factor in (\ref{3.6}) depending on $J$ is of the form
\beq{3.7}
\left.\frac{|h_J|}{\sqrt{g_J}}\right|_{J=I+1/2}=
\left(\frac{2I-1}{2I+3}\right)^{1/2}
\left.\frac{|h_J|}{\sqrt{g_J}}\right|_{J=I-1/2},\qquad
\left.\frac{|h_J|}{\sqrt{g_J}}\right|_{J=I-1/2}=
\left(\frac{I(2I-1)}{(I+1)(2I+1)}\right)^{1/2}.
\eeq
Therefore, the average effect for spin $J=I+1/2$ is less than the one for spin $J=I-1/2$.

To verify the statistical approach one can use the measurements of the deformation effect (\ref{2.15.2}) in p wave resonances. Indeed, averaging over the resonances with definite spin $J$ we obtain
\beq{3.13}
\av{p^J_D}=\frac{e_J}{4g_J}.
\eeq
For any $J$ this quantity can be evaluated with the use of (\ref{2.8}) and (\ref{2.10}). However, it is of interest that after averaging over all resonances the effect vanishes,
\beq{3.14}
\av{p_D}=\sum_{J=I\pm 1/2, 3/2}w(J)\,\av{p^J_D}=0.
\eeq
On the other hand, the root mean square takes the form
\beq{3.15}
\bar p_D=\left(\sum_J w(J)\,\av{\,(p^J_D)^2\,}\right)^{1/2}=
\left(\sum_J \frac{h^2_J+3e^2_J/4}{8g_J}\right)^{1/2}.
\eeq
The quantity $\bar p_D$ is defined only by the spin $I$ of the target nuclei. Results of its numerical estimation for some $I$ are presented in Table~\ref{t0}. We see that after measurements of the deformation effects in all known p wave resonances of the nucleus, one can get the average values $\av{p_D}$ and $\bar p_D$ and compare them with calculated quantities. Such a test would allow verifying the statistical approach proposed in this section.

\begin{table}
\caption{Root mean square deformation effect $\bar p_D$ as function of spin $I$ of the target nuclei.}
\label{t0}
\begin{center}
\begin{tabular}{|c|c|c|c|c|c|c|c|c|}
\hline
$I$ & 3/2 & 2 & 5/2 & 3 & 7/2 & 4 & 9/2 & 5\\
\hline
$\bar p_D$ & 0.296 & 0.354 & 0.395 & 0.427 & 0.452 & 0.472 & 0.489 & 0.502 \\
\hline
\end{tabular}
\end{center}
\end{table}

\section{Root mean square matrix element for PCTV in\-te\-rac\-tion}

In this section we discuss the magnitude of the mean matrix element, $v_T$, of the hypothetical PCTV interaction between the wave functions of the compound states. We assume that mean values of matrix elements between compound state functions of strong, $v$, P violating, $v_P$, and P conserving T violating, $v_T$, interactions have the same relation as the mean single particle matrix elements, $\av{V}_{s.p.}$, $\av{V_{PT}}_{s.p.}$ and $\av{V_{PCTV}}_{s.p.}$, of those interactions. Then to obtain a bound on $\lambda_T$ (\ref{1.1}) better than $10^{-4}$, the measurement of the five-fold correlation should be sensitive to the following value of $v_T$:
\beq{3.8}
v_T<10^{-4}\,v.
\eeq
Note that the matrix elements of weak and strong interactions are related by
\beq{3.9}
v_P\sim Gm_{\pi}^2\,v\sim 10^{-7}\,v,
\eeq
where $G$ is the Fermi constant and $m_{\pi}$ is the pion mass. Comparing (\ref{3.8}) and (\ref{3.9}), we obtain
\beq{3.10}
v_T<10^3\,v_P.
\eeq

Investigations of P violation in compound resonances have shown that $v_P\sim 1$~meV \cite{Mit01}. Thus, the proposed experiment on the measurement of five-fold correlation should be sensitive to the PCTV matrix element in the range
\beq{3.11}
v_T<1~\mbox{eV}.
\eeq

This bound on the matrix element of the PCTV interaction between the compound state wave functions results from the bound on the single particle matrix element $\av{V_{PCTV}}_{s.p.}$. There are also additional arguments for (\ref{3.11}). Let us assume that $v_T\sim 1$~eV. Since the average distance between p wave resonances in medium and heavy nuclei ($A\ge 100$) is equal to $D_p\sim 10$~eV, then the mixing may be on the scale of $v_T/D_p\sim 10^{-1}$. It has been noted in \cite{Bar93} that such strong mixing can be found in neutron-nucleus interaction without neutron polarization and nuclear alignment. Namely, the asymmetry was considered for $\gamma$ quanta emission after neutron capture along and opposite neutron momentum (asymmetry "forward-backward"). This asymmetry as function of neutron energy $E$ passes through zero in the p wave resonance. It has been shown that the deviation $\Delta E=E_0-E_p$ of the zero-energy $E_0$ from the energy $E_p$ of the resonance is sensitive to possible PCTV interaction. In \cite{Bar93} the results of measurements of forward-backward asymmetry of $\gamma$ quanta emission in the p wave resonance $E_p=7.0$~eV of the $^{113}$Cd nucleus were treated, and the evidence for $v_T/D_p<10^{-1}$, i.e. $v_T<1$~eV, was obtained.

It is worth noting that p wave resonances separated by a distance on the scale of 1~eV are not rare. For example, there are at least two pairs of such p wave resonances for the $^{127}$I nucleus, with the energies 52.20~eV and 53.82~eV, and the energies 352.0~eV and 353.3~eV (see \cite{Mit01}). Should the matrix element $v_T$ be on the scale of $\sim 1$~eV , then the mixing of the resonances in such pairs cannot be calculated in the framework of perturbation theory, i.e. formulae (\ref{2.22}), (\ref{2.23}) and (\ref{2.26}) were unsuitable (it is clear, however, that for such mixing, the PCTV effect in the resonances could be on the scale of unity).

Based on the current data it seems reasonable to expect that the mean matrix element $v_T$ is significantly less than 1~eV (if it is non-zero at all). Thus a new generation of experiments on the search for the hypothetical PCTV interaction should be focused on the region
\beq{3.12}
v_T\le 100~\mbox{meV}.
\eeq
A similar conclusion was reached in \cite{Dav99} from another basis: the bound 50~meV for $v_T$ was considered to be the goal for future measurements.

One usually discusses the bound on the PCTV interaction in terms of the ratio $\bar g_{\rho}$ of the T odd to T even $\rho$-meson exchange coupling constants. This $\rho$-meson exchange PCTV interaction violates charge symmetry (C). It is of interest that the bound $\bar g_{\rho} < 7\cdot 10^{-3}$ \cite{Sim97} extracted from the C violating effect in neutron--proton interactions exceeds the best one $\bar g_{\rho} < 6\cdot 10^{-2}$ \cite{Huf97} extracted from T violating effects in nuclear reactions (both bounds correspond to a 95\% confidence level). The estimation 50~meV for $v_T$ was obtained in \cite{Dav99} with the use of the bound on $\bar g_{\rho}$ from \cite{Sim97}. Note, however, that there is no reliable way of relating $\bar g_{\rho}$ with $v_T$ due to the complex structure of the compound state wave functions. On the other hand, the "quasitensor" nature of the PCTV interaction may result in the suppression of $v_T$ \cite{Gud91,Gud92}.

\section{Prospects of producing aligned targets}

To investigate T violation with the use of five-fold correlation one needs aligned nuclei (with spin $I\ge 1$) with low lying p wave resonances. Among the 20 nuclei studied with respect to P violation in the p wave resonances \cite{Mit01}, there are 10 nuclei with spin $I\ge 1$. They are listed in Table~\ref{t1} (the number of p wave resonances in the energy region 0-350 eV is presented).

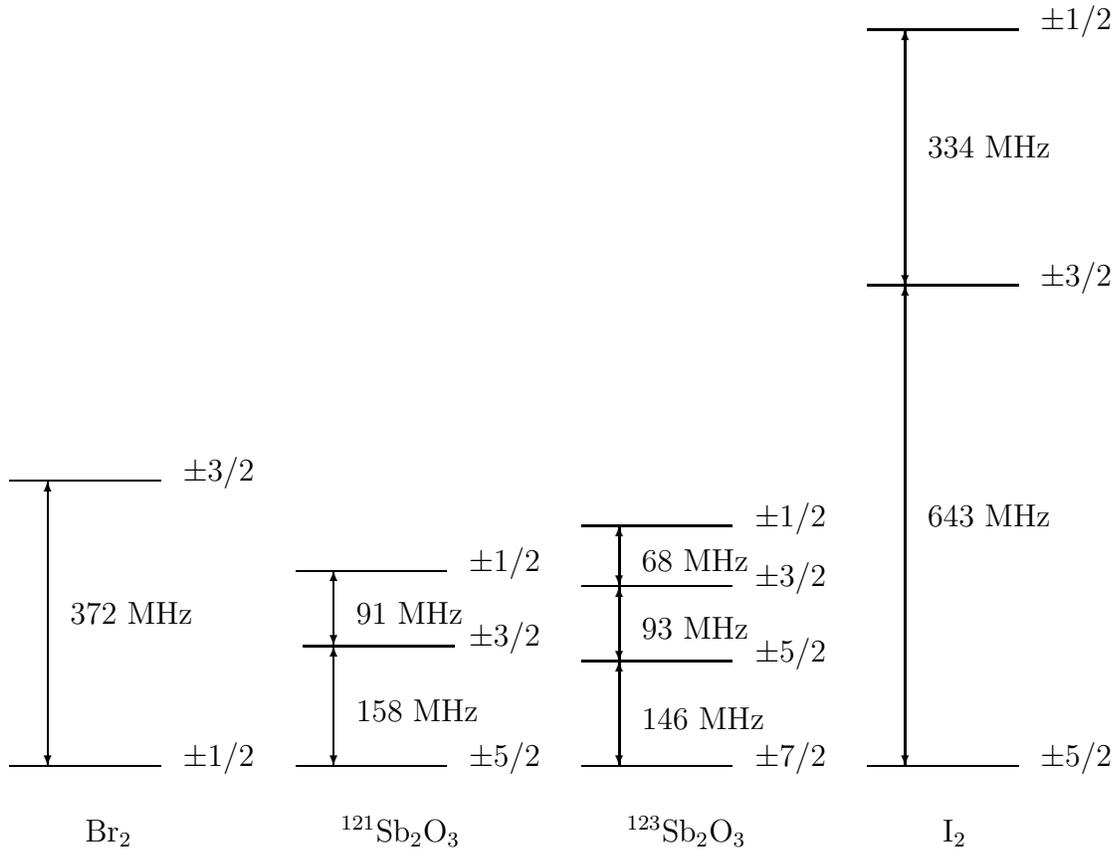
\begin{figure}
\begin{picture}(150,130)(0,-20)

\put(0,0){\line(1,0){20}}
\put(23,0){$\pm 1/2$}
\put(0,38){\line(1,0){20}}
\put(23,38){$\pm 3/2$}
\put(5,19){\vector(0,-1){19}}
\put(5,19){\vector(0,1){19}}
\put(8,19){372 MHz}
\put(10,-10){Br$_2$}

\put(38,0){\line(1,0){20}}
\put(61,0){$\pm 5/2$}
\put(39,16){\line(1,0){20}}
\put(61,16){$\pm 3/2$}
\put(38,26){\line(1,0){20}}
\put(61,26){$\pm 1/2$}
\put(43,8){\vector(0,-1){8}}
\put(43,8){\vector(0,1){8}}
\put(46,6){158 MHz}
\put(43,21){\vector(0,-1){5}}
\put(43,21){\vector(0,1){5}}
\put(46,19){91 MHz}
\put(44,-10){$^{121}$Sb$_2$O$_3$}

\put(76,0){\line(1,0){20}}
\put(99,0){$\pm 7/2$}
\put(76,14){\line(1,0){20}}
\put(99,14){$\pm 5/2$}
\put(76,24){\line(1,0){20}}
\put(99,24){$\pm 3/2$}
\put(76,32){\line(1,0){20}}
\put(99,32){$\pm 1/2$}
\put(81,7){\vector(0,-1){7}}
\put(81,7){\vector(0,1){7}}
\put(84,5){146 MHz}
\put(81,19){\vector(0,-1){5}}
\put(81,19){\vector(0,1){5}}
\put(84,17){93 MHz}
\put(81,28){\vector(0,-1){4}}
\put(81,28){\vector(0,1){4}}
\put(84,26){68 MHz}
\put(82,-10){$^{123}$Sb$_2$O$_3$}

\put(114,0){\line(1,0){20}}
\put(137,0){$\pm 5/2$}
\put(114,64){\line(1,0){20}}
\put(137,64){$\pm 3/2$}
\put(114,98){\line(1,0){20}}
\put(137,98){$\pm 1/2$}
\put(119,32){\vector(0,-1){32}}
\put(119,32){\vector(0,1){32}}
\put(122,32){643 MHz}
\put(119,81){\vector(0,-1){17}}
\put(119,81){\vector(0,1){17}}
\put(122,81){334 MHz}
\put(124,-10){I$_2$}

\end{picture}
\caption{\label{f1} Splitting of sublevels of the nuclei $^{81}$Br, $^{121}$Sb, $^{123}$Sb and $^{127}$I in the compounds Br$_2$, Sb$_2$O$_3$ and I$_2$ due to quadrupole interaction.}
\end{figure}

\begin{table}
\caption{Nuclei from \cite{Mit01} with spin $I\ge 1$.}
\label{t1}
\begin{center}
\begin{tabular}{|c|l|c|c|c|}
\hline
\parbox{1.0cm}{\centerline{N}} &
\parbox{2.0cm}{Isotope} &
\parbox{2.0cm}{\strut Natural\\ \strut abundance\\ \strut (\%)} &
\parbox{2.0cm}{\centerline{Spin}} &
\parbox{2.0cm}{\strut Number\\ \strut of p-wave\\ \strut resonances}\\
\hline
\strut 1 & $^{81}$Br & 49 & 3/2 & 1\\
2 & $^{93}$Nb & 100 & 9/2 & 18\\
3 & $^{105}$Pd & 22 & 5/2 & 29\\
4 & $^{115}$In & 96 & 9/2 & 40\\
5 & $^{121}$Sb & 57 & 5/2 & 17\\
6 & $^{123}$Sb & 43 & 7/2 & 6\\
7 & $^{127}$I & 100 & 5/2 & 20\\
8 & $^{131}$Xe & 21 & 3/2 & 1\\
9 & $^{133}$Cs & 100 & 7/2 & 28\\
10 & $^{139}$La & 100 & 7/2 & 1\\
\hline
\end{tabular}
\end{center}
\end{table}

\begin{figure}
\begin{center}
\mbox{\includegraphics*[scale=0.6]{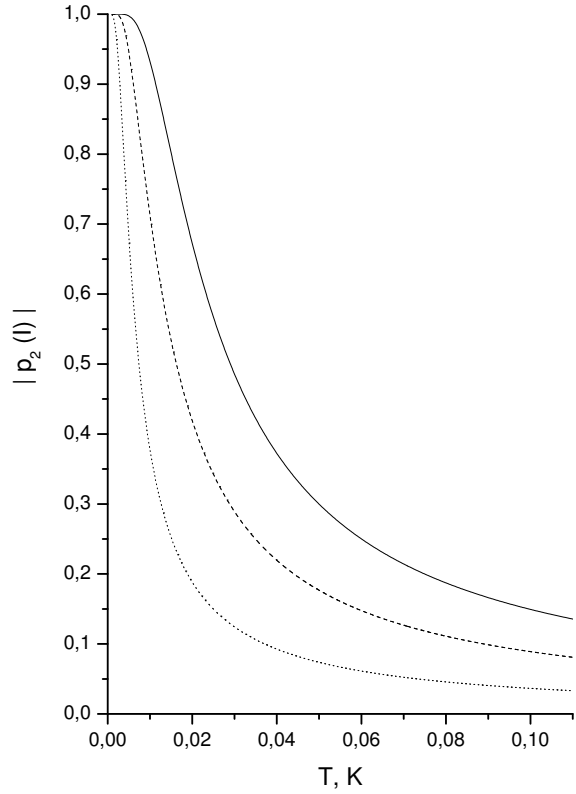}}
\caption{\label{f2} Modulus of the alignment parameter $p_2(I)$ as a function of the target temperature T calculated with the use of the data, presented in Fig.~1. Solid line -- nuclei $^{127}I$ in the crystal I$_2$, dashed line -- nuclei $^{81}$Br in the crystal Br$_2$, dotted line -- nuclei $^{121}$Sb and $^{123}$Sb in the crystal Sb$_2$O$_3$ (the parameters of alignment for both nuclei practically coincide for all temperatures).}
\end{center}
\end{figure}

\begin{table}
\caption{Results of the estimation of the TV effect for p wave resonances of $^{121}$Sb.}
\label{t2}
\begin{center}
\begin{tabular}{|c|c|c|c|c|}
\hline
\parbox{1.0cm}{\centerline{$\mu$}} &
\parbox{2.0cm}{\strut \centerline{$E^J_{\mu}$}} &
\parbox{2.0cm}{\strut \centerline{$g_J\Gamma^J_{n\mu}$, meV}} &
\parbox{2.0cm}{\centerline{$\Delta p_P\cdot 10^2$}} &
\parbox{2.0cm}{\strut \centerline{$\bar p_T$}}\\
\hline
1 & 37.9 & 0.0085 & 0.016 & 0.011 \\
\hline
2 & 55.21 & 0.014 & 0.018 & 0.008 \\
\hline
3 & 92.10 & 0.017 & 0.05 & 0.011 \\
\hline
4 & 110.7 & 0.037 & 0.06 & 0.007 \\
\hline
5 & 141.2 & 0.0081 & 0.12 & 0.018 \\
\hline
6 & 174.5 & 0.0054 & 0.033 & 0.133 \\
\hline
7 & 176.9 & 0.043 & 0.07 & 0.028 \\
\hline
8 & 184.7 & 0.126 & 0.033 & 0.009 \\
\hline
9 & 200.3 & 0.0054 & 0.076 & 0.045 \\
\hline
10 & 228.6 & 0.044 & 0.075 & 0.022 \\
\hline
11 & 235.9 & 0.019 & 0.16 & 0.052 \\
\hline
12 & 245.9 & 0.235 & 0.045 & 0.027 \\
\hline
13 & 249.0 & 0.128 & 0.052 & 0.048 \\
\hline
14 & 261.6 & 0.181 & 0.036 & 0.029 \\
\hline
15 & 265.8 & 0.167 & 0.048 & 0.040 \\
\hline
16 & 270.0 & 0.208 & 0.046 & 0.032 \\
\hline
17 & 274.8 & 0.153 & 0.050 & 0.030 \\
\hline
\end{tabular}
\end{center}
\end{table}

\begin{table}
\caption{Results of the estimation of the TV effect for p wave resonances of $^{123}$Sb.}
\label{t3}
\begin{center}
\begin{tabular}{|c|c|c|c|c|}
\hline
\parbox{1.0cm}{\centerline{$\mu$}} &
\parbox{2.0cm}{\strut \centerline{$E^J_{\mu}$}} &
\parbox{2.0cm}{\strut \centerline{$g_J\Gamma^J_{n\mu}$, meV}} &
\parbox{2.0cm}{\centerline{$\Delta p_P\cdot 10^2$}} &
\parbox{2.0cm}{\strut \centerline{$\bar p_T$}}\\
\hline
1 & 176.4 & 0.176 & 0.042 & 0.014 \\
\hline
2 & 186.1 & 0.154 & 0.035 & 0.019 \\
\hline

3 & 197.7 & 0.243 & 0.035 & 0.024 \\
\hline
4 & 202.0 & 0.160 & 0.12 & 0.034 \\
\hline
5 & 225.2 & 0.160 & 0.045 & 0.008 \\
\hline
\end{tabular}
\end{center}
\end{table}

\begin{table}
\caption{Results of the estimation of the TV effect for p wave resonances of $^{127}$I.}
\label{t4}
\begin{center}
\begin{tabular}{|c|c|c|c|c|}
\hline
\parbox{1.0cm}{\centerline{$\mu$}} &
\parbox{2.0cm}{\strut \centerline{$E^J_{\mu}$}} &
\parbox{2.0cm}{\strut \centerline{$g_J\Gamma^J_{n\mu}$, meV}} &
\parbox{2.0cm}{\centerline{$\Delta p_P\cdot 10^2$}} &
\parbox{2.0cm}{\strut \centerline{$\bar p_T$}}\\
\hline
1 & 7.51 & 0.00012 & 0.14 & 0.193 \\
\hline
2 & 10.34 & 0.0028 & 0.03 & 0.025 \\
\hline
3 & 13.94 & 0.0014 & 0.04 & 0.045 \\
\hline
4 & 24.63 & 0.00064 & 0.16 & 0.036 \\
\hline
5 & 52.20 & 0.00085 & 0.18 & 0.304 \\
\hline
6 & 53.82 & 0.019 & 0.02 & 0.016 \\
\hline
7 & 64.04 & 0.008 & 0.02 & 0.019 \\
\hline
8 & 85.84 & 0.0174 & 0.02 & 0.010 \\
\hline
9 & 101.1 & 0.014 & 0.03 & 0.013 \\
\hline
10 & 126.0 & 0.0021 & 0.16 & 0.071 \\
\hline
11 & 134.1 & 0.025 & 0.02 & 0.049 \\
\hline
12 & 136.9 & 0.040 & 0.016 & 0.033 \\
\hline
13 & 145.7 & 0.033 & 0.03 & 0.027 \\
\hline
14 & 153.6 & 0.096 & 0.02 & 0.009 \\
\hline
15 & 223.4 & 0.011 & 0.13 & 0.010 \\
\hline
16 & 256.8 & 0.052 & 0.04 & 0.005 \\
\hline
17 & 274.7 & 0.022 & 0.15 & 0.012 \\
\hline
18 & 282.1 & 0.0045 & 0.53 & 0.036 \\
\hline
19 & 352.0 & 0.088 & 0.064 & 0.080 \\
\hline
20 & 353.3 & 0.089 & 0.064 & 0.079 \\
\hline
\end{tabular}
\end{center}
\end{table}

Some time ago we proposed \cite{Ats00} a dynamic nuclear alignment (DNA) method to produce the targets needed for the study of five-fold correlation. This DNA method is analogous to the well known method of dynamic nuclear polarization but does not require an external magnetic field. So, there are no false effects connected with the precession of neutron spin resulting from both the magnetic field and pseudomagnetism, which takes place when one uses a polarized nuclear target. The DNA method can be used for the nuclei with large quadrupole splitting of spin sublevels in a single crystal.

The possibility of nuclear alignment was analyzed for nuclei from the Table~\ref{t1}, except $^{131}$Xe because the corresponding substance is gaseous (at room temperature). It appeared that the nuclei $^{93}$Nb, $^{105}$Pd, $^{115}$In, $^{133}$Cs and $^{139}$La could not be aligned either by the DNA or by the "brute force" method since they have very low values of quadrupole coupling constant in the studied compounds. As for the nuclei $^{81}$Br, $^{121}$Sb, $^{123}$Sb and $^{127}$I, they can be aligned, in principle, both by the "brute force" and DNA methods. The compounds LiIO$_3$, KBrO$_3$ and Sb$_2$O$_3$ are possible prospects in the last case. Realization of the DNA method is in progress for the compounds mentioned. The main problem is choosing the appropriate paramagnetic admixtures and their incorporation into the compounds.

On the other hand, we considered the prospects of the "brute force" method aligning the listed nuclei with rather high values of the quadrupole coupling constant. The preferable targets for the nuclei $^{81}$Br, $^{121,123}$Sb and $^{127}$I are the single crystals Br$_2$, Sb$_2$O$_3$ and I$_2$, respectively. In Fig.~\ref{f1} the splitting of nuclear sublevels for these targets is shown (the data are taken from \cite{Sem72}).
In Fig.~\ref{f2} the parameters of nuclear alignment as a function of target temperature are presented. It should be noted that over the last few years there has been impressive progress in the field of low temperature technique \cite{Uhl02}. Thus the construction of the target operating at a temperature of about 20 mK, which provides significant values for the parameter of nuclear alignment, is a realistic task.

\section{Results of the statistical estimation}

This section is devoted to the statistical estimation of the magnitude of five-fold correlation for the nuclei $^{121}$Sb, $^{123}$Sb and $^{127}$I, which can be aligned, for example, by the "brute force" method (see Fig.~2). No such estimation can be made for the nucleus $^{81}$Br, because only one p wave resonance is known for it. It is worth noting that according to \cite{Dav99}, in the case of null measurements one needs at least four resonances to put a substantial bound on the PCTV interaction.

\begin{figure}
\begin{center}
\mbox{\includegraphics*[scale=0.8]{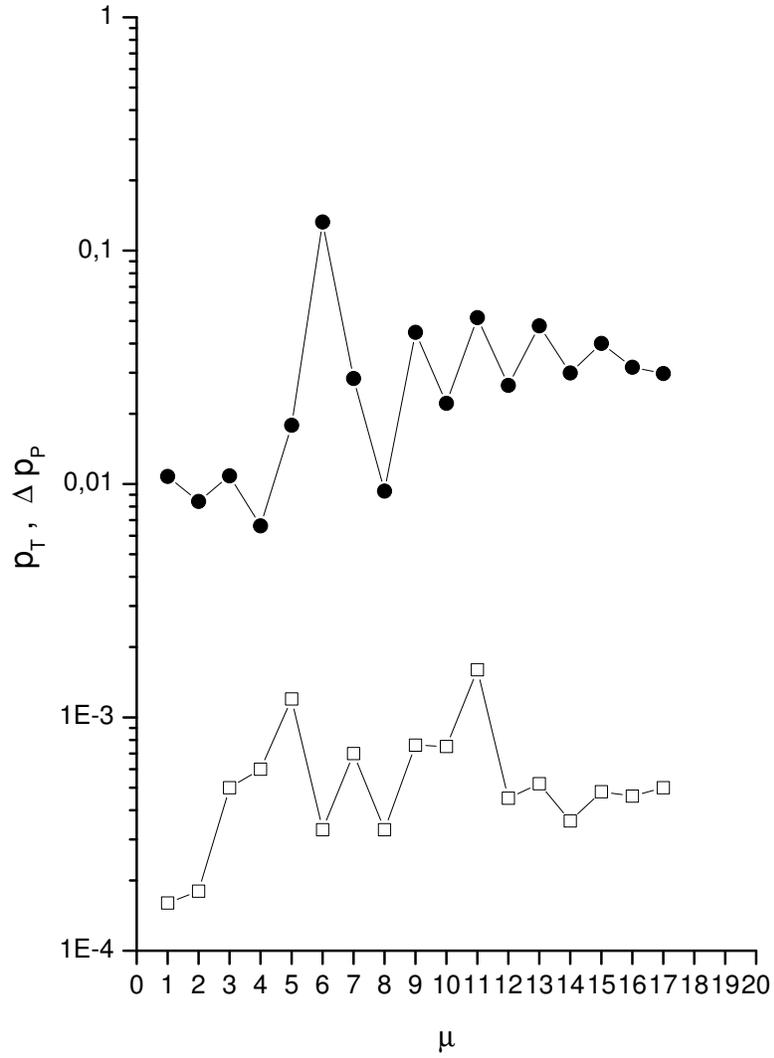}}
\caption{\label{f3} Comparison between statistical errors of the PV effect $\Delta p_P$ (open squares) and the evaluated PCTV effect $\bar p_T$ (dark circles) in p wave resonances labelled by $\mu$ for $^{121}$Sb nucleus (see Table~\ref{t2}).}
\end{center}
\end{figure}

\begin{figure}
\begin{center}
\mbox{\includegraphics*[scale=0.8]{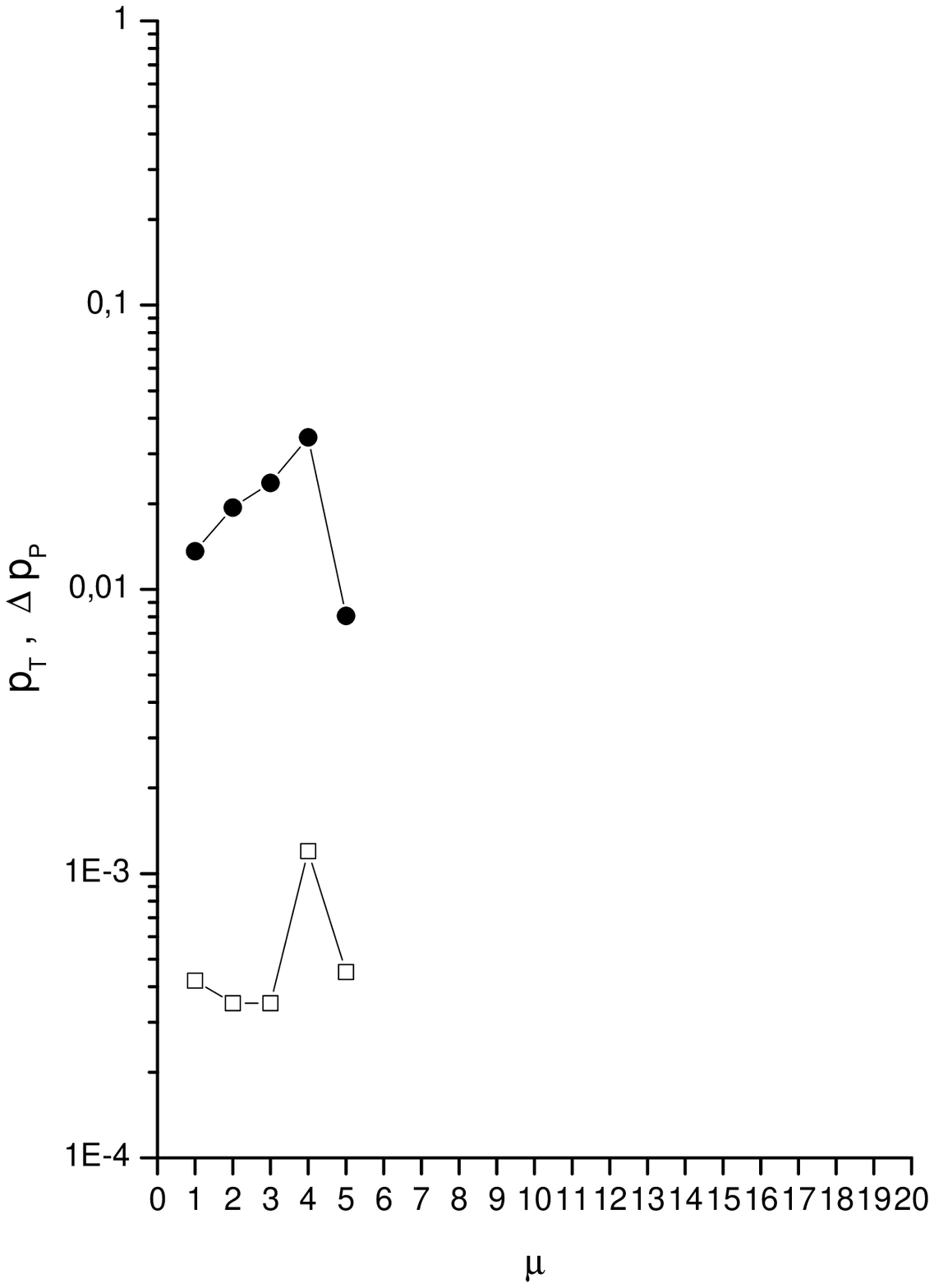}}
\caption{\label{f4} Comparison between statistical errors of the PV effect $\Delta p_P$ (open squares) and the evaluated PCTV effect $\bar p_T$ (dark circles) in p wave resonances labelled by $\mu$ for $^{123}$Sb nucleus (see Table~\ref{t3}).}
\end{center}
\end{figure}

\begin{figure}
\begin{center}
\mbox{\includegraphics*[scale=0.8]{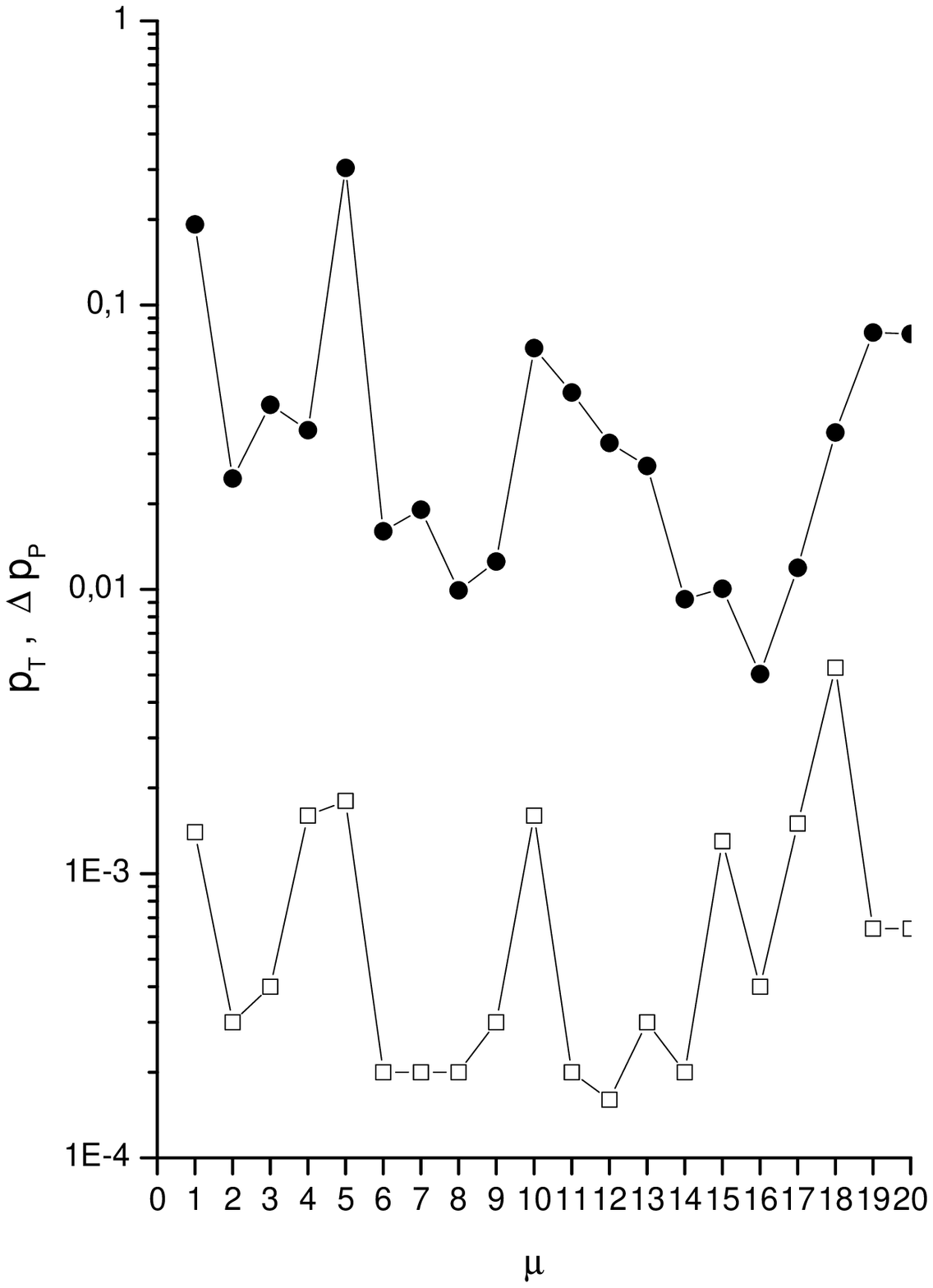}}
\caption{\label{f5} Comparison between statistical errors of the PV effect $\Delta p_P$ (open squares) and the evaluated PCTV effect $\bar p_T$ (dark circles) in p wave resonances labelled by $\mu$ for $^{127}$I nucleus (see Table~\ref{t4}).}
\end{center}
\end{figure}

The mean matrix element of the hypothetical PCTV interaction mixing p wave resonances is taken to be equal to $v_T=100$~meV (see (\ref{3.12})). The magnitude of the PCTV effect $\bar p_T$ in a p wave resonance is given by  Eq.~(\ref{3.6}). For the sake of definiteness we assume $J=I-1/2$ for the spin of the resonance, where $I$ is the spin of the target nucleus. The results of estimations are presented in Tables~\ref{t2}-\ref{t4}. In these Tables for each resonance with number $\mu$ the position $E^J_{\mu}$, parameter $g_J\Gamma^J_{n\mu}$ and the achieved statistical accuracy $\Delta p_P$ of the PV effect measurement are indicated, taken from \cite{Mit01}, as well as the calculated value of $\bar p_T$. A comparison between $\Delta p_P$ and $\bar p_T$ in a more clear form is presented in Figs.~\ref{f3}-\ref{f5}.

The surprising thing is the correlation between $\bar p_T$ and $\Delta p_P$. The reason is as follows. The magnitude of the statistical error $\Delta p_P$ is dominantly determined by the neutron width of the resonance: the more weak the resonance, the larger the error $\Delta p_P$ (see Tables~\ref{t2}-\ref{t4}). On the other hand, let us assume that near the weak resonance there exists a strong resonance mixing to the former. Then according to (\ref{3.6}), the PCTV effect in the weak resonance is enhanced by the factor $(\Gamma^J_{n\mu'}/\Gamma^J_{n\mu})^{1/2}$, where $\Gamma^J_{n\mu'}$ and $\Gamma^J_{n\mu}$ are the neutron widths of strong and weak resonances, respectively. This factor should be named the factor of kinematical enhancement by analogy with the similar factor of enhancement of PV effects in (\ref{2.20}) (recall that neutron widths of s wave resonances mixing to p wave ones far exceed the neutron widths of the p wave resonances).

For example, the predicted effect $\bar p_T$ is rather high in the 1st, 5th and 10th resonances of the $^{127}$I nucleus in correlation with large statistical errors $\Delta p_P$ in the same ones (see Table~\ref{t4} and Fig.~5). The common feature of these resonances is that they are all weak (with low values of $g_J\Gamma^J_{n\mu}$). On the other hand, their neighbours, the 2nd, 6th and 11th resonances, respectively, are relatively strong. Thus, a mixture of weak and strong resonances results in large $\bar p_T$ in the listed weak resonances.

Statistical estimation for root mean square deformation effects for p wave resonances of the nuclei $^{121}$Sb, $^{127}$I ($I=5/2$) and $^{123}$Sb ($I=7/2$) is presented in Table~\ref{t0}.

\section{Conclusions}

In this paper we discuss five-fold (P even, T odd) correlation in the interaction of slow polarized neutrons with aligned nuclei as a prospective test of time reversal invariance due to possible enhancement of T violating effects in compound resonances. Possible nuclear targets are considered which can be aligned both dynamically as well as by the "brute force" method at low temperature.

It is shown that the nuclei $^{121}$Sb, $^{123}$Sb and $^{127}$I are the most promising targets for five-fold correlation studies. Indeed, on the one hand, a lot of p wave resonances are known for these nuclei. On the other hand, quadrupole splitting of spin sublevels in the appropriate compounds is rather high. Thus significant values of nuclear alignment can be obtained even by the well known "brute force" method at temperature of about 20 mK.

Statistical estimation is performed for a magnitude of five-fold correlation for low lying p wave resonances of the $^{121}$Sb, $^{123}$Sb and $^{127}$I nuclei. It is shown that at the mean matrix element $v_T=100$~meV of the PCTV interaction mixing the resonances, the expected effects exceed the statistical errors of recently measured PV effects by one to two orders of magnitude. This means that in the proposed experiment, one can reach sensitivity on the scale of $v_T\sim 100$~meV, i.e. to lower the bound on $\lambda_T$ to the level of 10$^{-5}$ (at best, to discover PCTV interaction). An improvement of measurement conditions may be possible at new high intensity spallation neutron sources -- SNS (USA) and JSS (Japan).
\bigskip

{\small We are grateful to V.P.Alfimenkov and L.B.Pikelner for valuable discussions. The work is supported by grant NS-1885.2003.2 and RFBR grant 03-02-16050.}


\bigskip

\begin{thebibliography}{99}
\bibitem{Bun82} V. E. Bunakov, V. P. Gudkov, JETP Lett. 36 (1982) 328.
\bibitem{Bun88} V. E. Bunakov, Phys. Rev. Lett. 60 (1988) 2250.
\bibitem{Sus80} O. P. Suskov, V. V. Flambaum, JETP Lett. 32 (1980) 352.
\bibitem{Mit01} G. E. Mitchell et al., Phys. Rep. 354 (2001) 157.
\bibitem{Her96} P. Herczeg, in: Parity and Time Reversal Violation in Compound Nuclear
States and Related Topics, Eds. N. Auerbach and J. D. Bowman, World
Scientific, Singapure, 1996, p.214.
\bibitem{Kur01} A. Kurylov, G. C. McLaughlin, M. J. Ramsey-Musolf, Phys. Rev. D 63 (2001) 076007.
\bibitem{Gou02} C. R. Gould, in: Astrophysics, Symmetries, and Applied Physics at Spallation Neutron Sources (Proceedings of the Workshop, ORNL, March 10-12, 2002), Eds. P. E. Koehler, C. R. Gould, R. Haight, T. E. Valentine, World Scientific, Singapore, 2002, p.209.
\bibitem{Bar03} A. L. Barabanov, A. G. Beda, A. F. Volkov, in: Neutron Spectroscopy, Nuclear Structure, Related Topics (Proceedings of the 10th International Seminar on Interaction of Neutrons with Nuclei, Dubna, May 22-25, 2002), Dubna, JINR, 2003, p.26; Czech. J. Phys. Suppl. B 53 (2003) B371.
\bibitem{Bary83} V. G. Baryshevskii, Sov. J. Nucl. Phys. 38 (1983) 699.
\bibitem{Bar86} A. L. Barabanov, Sov. J. Nucl. Phys. 44 (1986) 755.
\bibitem{Kab86} P. K. Kabir, in: The Investigation of Fundamental Interactions with Cold Neutrons, Ed. G. L. Greene,  NBS Special Publication 711, Washington, 1986, p. 81.
\bibitem{Con93} H. E. Conzett, Phys. Rev. C 48 (1993) 423.
\bibitem{Con95} H. E. Conzett, Phys. Rev. C 52 (1995) 1041.
\bibitem{Gou90} C. R. Gould, D. G. Haase, N. R. Robertson, H. Postma, J. Bowman, Int. J. Mod. Phys. A 5 (1990) 2181.
\bibitem{Bun90} V. E. Bunakov, E. D. Davis, H. A. Weidenmuller, Phys. Rev. C 42 (1990) 1718.
\bibitem{Gud91} V. P. Gudkov, Nucl. Phys. A 524 (1991) 668.
\bibitem{Huf98} P. R. Huffman, C. R. Gould, D. G. Haase, J. Phys. G: Nucl. Phys. 24 (1998) 763.
\bibitem{Huf02} P. R. Huffman, in: Astrophysics, Symmetries, and Applied Physics at Spallation Neutron Sources (Proceedings of the Workshop, ORNL, March 10-12, 2002), Eds. P. E. Koehler, C. R. Gould, R. Haight, T. E. Valentine, World Scientific, Singapore, 2002, p.217.
\bibitem{Huf97} P. R. Huffman, N. R. Robertson, W. S. Wilburn, C. R. Gould, D. G. Haase, C. D. Keith, B. W. Raichle, M. L. Seely, J. R. Walston,  Phys. Rev. C 55 (1997) 2684.
\bibitem{Bar87} A. L. Barabanov, Sov. J. Nucl. Phys. 45 (1987) 597.
\bibitem{Bun81} V. E. Bunakov, V. P. Gudkov, Z. Phys. A 303 (1981) 285.
\bibitem{Bar93} A. L. Barabanov, E. I. Sharapov, V. R. Skoy, C. M. Frankle, Phys. Rev. Lett. 70 (1993) 1216.
\bibitem{Dav99} E. D. Davis, C. R. Gould, Phys. Lett. B 447 (1999) 209.
\bibitem{Sim97} M. Simonius, Phys. Rev. Lett. 78 (1997) 4161.
\bibitem{Gud92} V. P. Gudkov, Phys. Rep. 212 (1992) 77.
\bibitem{Ats00} V. A. Atsarkin, A. L. Barabanov, A. G. Beda, V. V. Novitsky, Nucl. Instr. Methods A 440 (2000) 626.
\bibitem{Sem72} G. K. Semin, T. A. Babushkina, G. G. Yakobson, Primenenie YaKR v khimii, Leningrad, Khimiya, 1972 (in Russian).
\bibitem{Uhl02} K. Uhlig, Cryogenics 42 (2002) 73.
\end{thebibliography}
\end{document}